\documentclass[10pt]{article}
\usepackage[margin=1.5in]{geometry}
\usepackage{graphicx}
\usepackage{amssymb}
\usepackage{amsmath}
\usepackage{tikz}
\usepackage{url}
\usepackage{color}
\usepackage{savetrees}
\usepackage{listings}
\usepackage{hyperref}
\usepackage{authblk}
\usetikzlibrary{shapes}

\title{A Simple Solution to the Level-Ancestor Problem}
\author[1]{Gaurav Menghani}
\author[2]{Dhruv Matani}
\affil[1]{Google AI, Mountain View (CA) - 94043}
\affil[2]{Facebook Inc., Menlo Park (CA) - 94025}
\date{\vspace{-5ex}}
\begin{document}
\maketitle
\vspace*{0.25cm}
\begin{abstract}
A Level Ancestory query LA($u$, $d$) asks for the the ancestor of the node $u$ at a depth $d$. We present a simple solution, which pre-processes the tree in $O(n)$ time with $O(n)$ extra space, and answers the queries in $O(\log\ {n})$ time. Though other optimal algorithms exist, this is a simple enough solution that could be taught and implemented easily.

\end{abstract}
\vspace*{0.25cm}

\section{Introduction}
The Level-Ancestor problem has been well-studied \cite{Bender2004, Berkman1994, Dietz1991}. \cite{Bender2004} provides a simplified algorithm, that combines two complementary approaches, the Jump Pointer algorithm (to jump half-way to the target height in a $O(1)$ hops), and the Ladder algorithm (to cover the remaining height using a single `ladder' in $O(1)$). However, while the algorithm is simpler and has a much better constant factor than previously published optimal algorithms such as \cite{Berkman1994}, it still requires several non-trivial book-keeping for the $O(n)$ preprocessing and $O(1)$ query complexity (jump-pointers, ladders, macro-micro tree, and the lookup tables).

While the presented method is equivalent in complexity to the ladder algorithm of \cite{Bender2004}, it is arguably simple enough for even a junior undergraduate to implement.

\section{Our Method}

\subsection{Preprocessing}
We perform a \textit{pre-order traversal} of the tree we are
interested in. We keep a counter $\mathtt{c}$ initialized to 0. 

Each time we see a new node $\mathtt{n}$, during the pre-order traversal:
\begin{enumerate}
    \item Increment $\mathtt{c}$.
    \item Assign $\mathtt{c}$ as the \textit{label} of $\mathtt{n}$.
    \item Append $\mathtt{c}$ to the dynamic array associated with the level of $\mathtt{n}$.
    \item Insert a pointer to $\mathtt{n}$ in the hash table with $\mathtt{c}$ as the key.
\end{enumerate}

It is easy to see that all the four steps are $O(1)$ in time. Since they are invoked exactly once for each of the $n$ nodes of the tree in its pre-order traversal, the overall complexity of this preprocessing is $O(n)$.

\begin{figure}[t]
\includegraphics[scale=0.125]{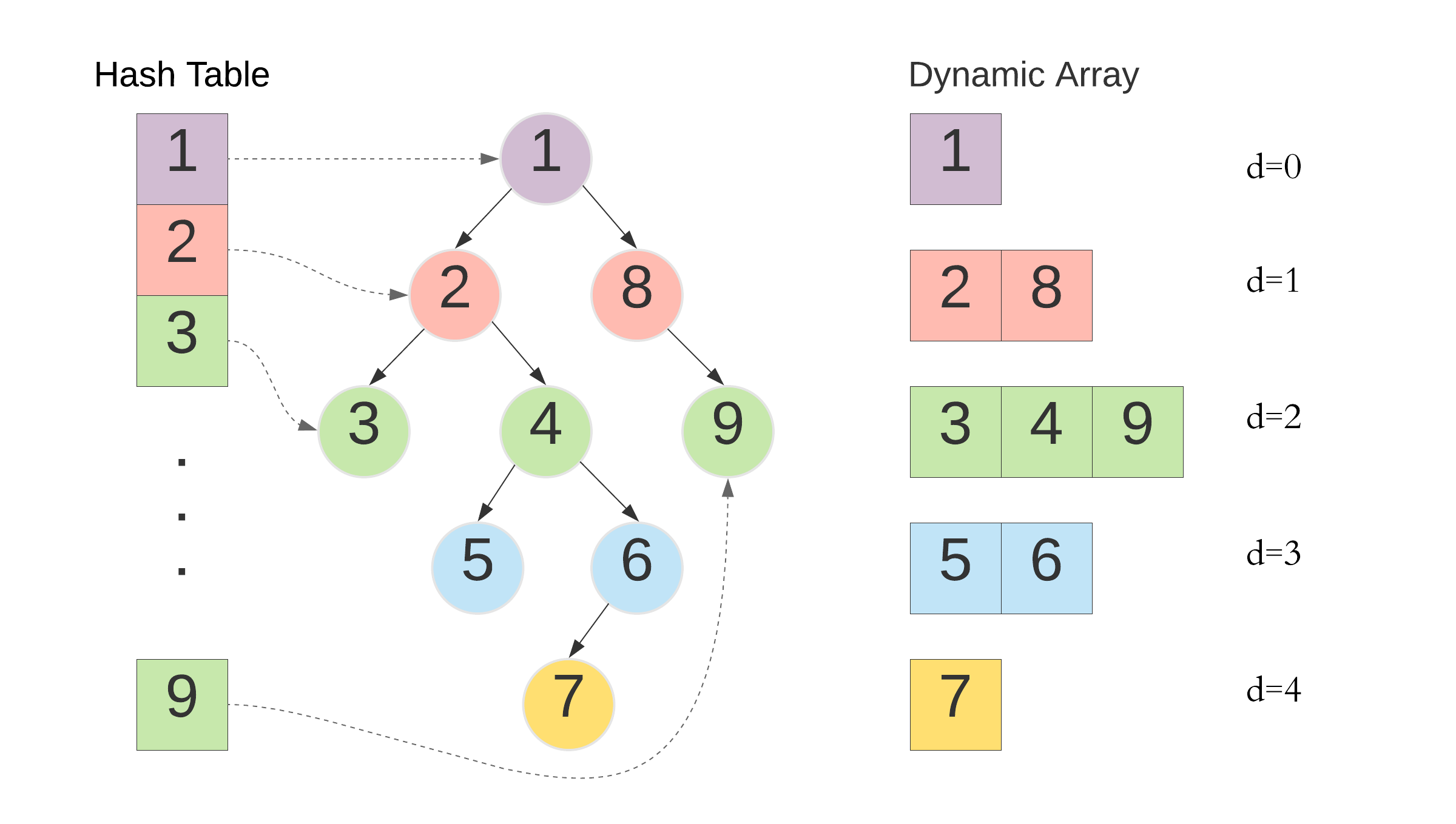}
\centering
\caption{Example of the preprocessing of the tree. Each node has a label assigned during the pre-order traversal. The left hand side shows the hash table which lets the user find the node, given this label. The right hand side shows the dynamic arrays at each level.}
\end{figure}

\subsection{Observation}
The dynamic arrays mentioned in step 3 of the preprocessing will always have the labels in a strictly increasing order. Since we process the nodes in a preorder traversal, when adding a node $u$, all its ancestors would have already been added into their respective dynamic arrays. If $u$ is the latest node being processed, all ancestors of $u$ would at that time have the largest labels in their respective arrays, but smaller than the label of $u$, since it is the latest node being processed.

By induction, given any node $v$, its ancestor at depth $d$ would have the largest label at that depth, but smaller than the label of $v$.

\subsection{Query}

To answer the query LA($v$, $d$):
\begin{enumerate}
    \item Retrieve the label $\mathtt{l}$ associated with the node $v$.
    \item Look up the largest element $\mathtt{s}$ smaller than $\mathtt{l}$ in the dynamic array associated with the query depth $d$.
    \item Return the pointer in the hash table for the key $\mathtt{s}$.
\end{enumerate}

It is easy to see both steps 1 and 3 are $O(1)$.

Step 2 works because of the observations in section 2.2. Since the dynamic arrays have the labels in strictly increasing order, we can find the largest element smaller than the label $\mathtt{l}$ in $O(\log\ {n})$ time using binary search. Thus the overall time complexity of the query is $O(\log\ {n})$.

It is further possible to reduce it by a constant factor to $O(\log\ {\sqrt{n}})$ time using \cite{DhruvFastBinarySearch}.

\subsection{Summary}
We suggest an $O(n)$ preprocessing, and $O(\log\ {n})$ query algorithm for the Level Ancestor problem. This is a simple to implement and understand algorithm, with a reasonable space, time, and implementation complexity.

\end{document}